\documentclass[sn-mathphys]{sn-jnl}% Math and Physical Sciences Reference Style
\usepackage{multirow}%
\usepackage{amsmath,amssymb,amsfonts,bm}%
\usepackage{amsthm}%
\usepackage{mathrsfs}%
\usepackage{graphicx}
\numberwithin{equation}{section}

\begin{document}

\title[Article Title]{Evolution of vortex filaments and reconnections in the Gross-Pitaevski equation and its approximation by the binormal flow equation.}

\author[1]{\fnm{M.} \sur{Array\'as}}%\email{manuel.arrayas@urjc.es}
\author[2]{\fnm{M. A.} \sur{Fontelos}}%\email{marco.fontelos@icmat.es}
\author[3]{\fnm{M.d.M.} \sur{Gonz\'alez}}%\email{mariamar.gonzalezn@uam.es}
\author[1]{\fnm{C.} \sur{Uriarte}}%\email{carlos.ugonzalez@urjc.es}

\affil[1]{\orgdiv{Área de Electromagnetismo}, \orgname{Universidad Rey Juan Carlos}, \orgaddress{\street{Tulipán s/n}, \city{Mostoles}, \postcode{28933}, \state{Madrid}, \country{Spain}}}

\affil[2]{\orgdiv{Instituto de Ciencias Matem\'aticas}, \orgname{CSIC}, \orgaddress{\street{Nicolás Cabrera, 13-15}, \city{Campus de Cantoblanco}, \postcode{28049}, \state{Madrid}, \country{Spain}}}

\affil[3]{\orgdiv{Departamento de Matem\'aticas}, \orgname{Universidad Aut\'onoma de Madrid}, \orgaddress{\street{Francisco Tom\'as y Valiente, 7}, \city{Campus de Cantoblanco}, \postcode{28049}, \state{Madrid}, \country{Spain}}}


\abstract{The evolution of a vortex line following the binormal flow equation (i.e. with
a velocity proportional to the local curvature in the direction of the binormal
vector) has been postulated as an approximation for the evolution of vortex
filaments in both the Euler system for inviscid incompressible fluids and the
Gross-Pitaevski equation in superfluids. We address the issue of whether this
is a suitable approximation or not and its degree of validity by using rigorous
mathematical methods and direct numerical simulations. More specifically, we
show that, as the vortex core thickness goes to zero, the vortex core moves (at
leading order and for long periods of time) with a velocity proportional to its local
curvature and the binormal vector to the curve. The main idea of our analysis lies
in a reformulation of the Gross-Pitaevski equation in terms of associated velocity
and vorticity fields that resemble the Euler system written in terms of vorticity
in its weak form. We also present full numerical simulations aimed to compare
Gross-Pitaevski and binormal flow in various physical situations of interest such
as the periodic evolution of deformed vortex rings and the reconnection of vortex
filaments.}

%\keywords{keyword1, Keyword2, Keyword3, Keyword4}



\maketitle

\section{Introduction}\label{sec1}
As in the case of the classical turbulence (see \cite{Batchelor} for a general reference), the turbulent regime in a superfluid is not fully understood. The superfluid flow is influenced by quantum effects and the visualisation techniques available for classical fluids are not directly applicable to the quantum case \cite{Vinen02}.

The pioneering work by Schwarz \cite{Schwarz85, Schwarz88} investigated quantum turbulence using a vortex filament dynamics model based on an approximation of the familiar Biot-Savart expression, named as LIA (from localised induction approximation), coupled to an external frictional force. This approach has been improved since then in subsequent works \cite{ATN}.  As one example we should mention the study of steady-sate counterflow quantum turbulence \cite{Adachi10}, where the LIA approximation was extended. In the LIA approximation, the evolution of a vortex line follows the binormal flow equation (i.e. with a velocity proportional to the local curvature in the direction of the binormal vector)

In this study, we first deduce the binormal flow dynamics as the leading order contribution to the evolution of the filament core over long time intervals. Specifically, we demonstrate that, as the vortex core thickness $\epsilon$ approaches zero, the vortex core moves with a velocity given by $\mathbf{v} = \lvert \log \epsilon \rvert \kappa \mathbf{b}$, where $\kappa$ is the local curvature and $\mathbf{b}$ is the binormal vector to the vortex core. This mathematical approached is based on a reformulation of the Gross-Pitaevski equation in terms of velocity and vorticity fields, which closely resembles the weak form of the Euler system written in terms of vorticity. This is in contrast to previous formal approaches that exploited the facts that Gross-Pitaevski equation can be transformed into a compressible Euler system by means of Madelung transformation, and that Euler system formally supports solutions in the form of vortex filaments.

Second, we provide strong numerical evidence of the accuracy of this approximation by means of two classical scenarios: the evolution of a perturbed vortex ring leading to oscillations, and the reconnection of two vortex filaments leading to singular corners whose later evolution possesses precise self-similar features.

We will write the Gross-Pitaevskii (GP) equation to describe the condensate as \cite{Gross61,Pitaevskii61},
\begin{equation*}
  i\hbar \frac{\partial \phi}{\partial t}= -\frac{\hbar^2}{2m}\nabla^2\phi + (V + g \lvert \phi \rvert^2 -\mu)\phi.
  \label{GP}
\end{equation*}
Here $V$ denotes any external potential,  $g=4\pi a_s\hbar^2/m$ the coupling parameter,
with $a_s$ being the $s$-wave scattering length of the atoms, and $\mu$ the chemical potential.
For $^4$He, at zero temperature, $\mu \sim 7.15 k_B$ \cite{Syvokon}.

We introduce a convenient rescaling of variables,
\begin{equation*}
  % \label{eq:dim}
  x = \frac{\hbar}{\sqrt{2m \mu \epsilon^2}}\tilde{x},\,\,\, t = \frac{\hbar}{\mu \epsilon^2}\tilde{t},\,\,\, \phi = \sqrt{\frac{\mu}{g} }\,u,
\end{equation*}
being $\epsilon$ a dimensionless small quantity that we will take as an expansion parameter. Then, without  any external potential ($V=0$), the GP equation can be cast into the form
\begin{equation}
  \label{eq:GP}
      i u_t= -\Delta u + \frac{1}{\epsilon^2}(|u|^2 -1)u,
\end{equation}
where the tildes in the dimensionless quantities has been dropped and $\Delta\equiv\nabla^2$. This is the GP model that we will use in the following sections.

\section{Euler equation from Gross-Pitaevskii}
In order to reformulate the GP equation \eqref{eq:GP},  let us introduce
the product
\begin{equation*}
  \label{eq:product}
  (a,b) = \frac{1}{2} (\overline{a}b + a\overline{b}),
\end{equation*}
where $\overline{a}$ denotes the complex conjugate of $a$. Now we take the time derivative of the product $(iu,\nabla u)$,
\begin{eqnarray*}
\frac{\partial }{\partial t}(iu,\nabla u) &=&(iu_{t},\nabla u)-(u,\nabla
iu_{t})
\end{eqnarray*}%
and we make use of the GP \eqref{eq:GP} to substitute $iu_t$ on the right hand side. Using the fact that
\[
(u,\nabla \Delta u)=\nabla(u,\Delta u) -(\nabla u,\Delta u)
\]%
and%
\begin{eqnarray*}
&&((\left\vert u\right\vert ^{2}-1)u,\nabla u)-(u,\nabla ((\left\vert
u\right\vert ^{2}-1)u)) = -\nabla \frac{\left\vert u\right\vert ^{4}}{2},
\end{eqnarray*}%
we conclude
\begin{equation*}
\frac{\partial }{\partial t}(iu,\nabla u)=-2(\Delta u,\nabla
u)+\frac{1}{\epsilon^2}\nabla \left((\Delta u,u)-\frac{\left\vert u\right\vert
    ^{4}}{2}\right)
\label{eq:GP2}
\end{equation*}
or, taking the $\nabla \times$ operator%
\begin{equation}
-\frac{\partial }{\partial t}\left( \nabla \times (iu,\nabla u)\right)
=2\nabla \times (\Delta u,\nabla u).
\label{eq:GPEu}
\end{equation}

Our aim is to show that this equation behaves asymptotically as Euler's
equation for a thin vortex filament when $\epsilon \ll 1$. Let us write
\[
  u=\rho\, e^{i\psi},
\]
so that%
\[
(\Delta u,\nabla u)=\nabla \rho \left( \Delta \rho -\rho \left\vert \nabla
\psi \right\vert ^{2}\right) +2\rho \nabla \psi \left( \nabla \rho \cdot
\nabla \psi \right) +\rho ^{2}\Delta \psi \nabla \psi.
\]%

Now if we take $\rho(r)\approx 1$ outside a core of order $\epsilon$, and $\psi = \theta$ being the angular coordinate in a normal plane of the core region, we get from \eqref{eq:GPEu}
\begin{equation}
-\frac{\partial }{\partial t}\left( \nabla \times (\rho ^{2}\nabla \psi
)\right) =2\nabla \times \left( \rho ^{2}\Delta \psi \nabla
\psi \right) +...  \label{a1}
\end{equation}%
where we have neglected terms at the right hand side that will be small
compared with the main term. Note that, for a filament and in terms of the
local coordinates $x^{\prime }=(x_{1},x_{2},s)$ we have, at leading order,
\[
\Delta \psi =\Delta _{x^{\prime }}\psi -\kappa \nabla \psi \cdot \mathbf{n}+...
\]%
Moreover, in terms of the Frenet-Serret trihedron,%
\[
\nabla \psi =(\nabla \psi \cdot \mathbf{n})\mathbf{n}+(\nabla \psi \cdot \mathbf{b})\mathbf{b}+\frac{d\psi }{ds}\mathbf{t},
\]%
so that%
\begin{equation}
\rho ^{2}\Delta \psi \nabla \psi =\rho ^{2}\Delta _{x^{\prime }}\psi \nabla
\psi -\kappa \rho ^{2}\left\vert \nabla \psi \cdot \mathbf{n}\right\vert ^{2}\mathbf{n}+...
\label{a2}
\end{equation}
\\
We substitute this last expression \eqref{a2} into the equation \eqref{a1}.
Doing that, let us fix our attention to the second term at the right hand side and note%
\[
\nabla \times (\varphi \mathbf{F})=\nabla \varphi \times \mathbf{F}+\varphi
\nabla \times \mathbf{F}.
\]%
Thus, multiplying by a time-independent test function $\varphi $ and
integrating over the volume (in the spirit of \cite{Jerard}, where a rigorous mathematical approach to the problem is considered) we get%
\[
2\int \nabla \times (-\kappa \rho ^{2}\left\vert \nabla \psi \cdot
\mathbf{n}\right\vert ^{2}\mathbf{n})\varphi\,d^3{\bf r} =2\int \kappa \rho ^{2}\left\vert \nabla \psi
\cdot \mathbf{n}\right\vert ^{2}\mathbf{n}\times \nabla \varphi\,d^3{\bf r}.
\]%
By writing%
\[
\nabla \varphi =\varphi _{n}\mathbf{n}+\varphi _{b}\mathbf{b}+\varphi _{s}\mathbf{t},
\]%
we can substitute $\mathbf{n}\times \nabla \varphi$, thus
\begin{equation}
2\int \kappa \rho ^{2}\left\vert \nabla \psi \cdot \mathbf{n}\right\vert ^{2}\left(
\varphi_b \mathbf{t}-\varphi _{s}\mathbf{b}\right)\,d^3{\bf r} \sim 2\pi
\left\vert \log \epsilon \right\vert \left[ \int \kappa \left( \mathbf{b}\cdot
\nabla \varphi \right) \mathbf{t}ds+\int \frac{d(\kappa \mathbf{b})}{ds}\varphi ds\right],
\label{b1}
\end{equation}%
where we have used, at each cross-section
\[
\int_{0}^{R/\epsilon }\rho ^{2}\left\vert \nabla \psi \cdot \mathbf{n}\right\vert
^{2}rdrd\theta =\pi \left\vert \log \epsilon \right\vert +...
\]%
On the other hand, looking at the left hand side of \eqref{a1} and approximating 
$$\nabla \times (\rho ^{2}\nabla \psi)= \nabla \times \left(\frac{\rho ^{2}}{r}{\bf e}_\theta \right)\sim\delta_\Gamma \mathbf t,$$ where $\delta_\Gamma$ is the Dirac delta function supported on the curve $\Gamma$, we have
\begin{equation}
\frac{\partial }{\partial t}\int \left( \nabla \times (\rho ^{2}\nabla \psi
)\right) \varphi\,d^3{\bf r} \sim 2\pi\frac{\partial }{\partial t}\int \varphi (x_{0}(s,t)%
\mathbf{t}\,ds=2\pi\int \left[ \left( v\cdot \nabla \varphi \right) \mathbf{t}%
+\varphi \frac{d\mathbf{t}}{dt}\right] ds  \label{b2}
\end{equation}%
and, keeping in mind that
\begin{equation}
\frac{d\mathbf{t}}{dt}=\frac{dv}{ds},
\label{tgvel}
\end{equation}
after comparing \eqref{b1} and \eqref{b2} we get
\begin{equation}
v=- \left\vert \log \epsilon \right\vert \kappa \mathbf{b},
\label{biflow}
\end{equation}
if $\psi=\theta$ or
$$v=\left\vert \log \epsilon \right\vert \kappa \mathbf{b},$$
if $\psi=-\theta$. When $\psi=n\theta$, $n=\pm 1,\pm 2,...$ we would have $v=\mp n\left\vert \log \epsilon \right\vert \kappa \mathbf{b}$.

We conclude that the velocity of the filament is in the binormal direction, so we get the result of a binormal flow approximation. We remark that the first term at the right hand side of \eqref{a2}
cancels due to symmetry when integrated in front of a test function.

We finish by linking our result with the better known Euler's equation that we write in
the form
\begin{equation*}
\frac{\partial \omega }{\partial t}+v\cdot \nabla \omega -\omega \cdot
\nabla v=0,
\label{euler}
\end{equation*}%
with
\[\nabla \cdot v=0,
\]
and%
\begin{equation}
\nabla \times v=\omega.   \label{c1}
\end{equation}%
If
\[
\omega =\nabla \times v_{\theta }=\delta _{\Gamma }\mathbf{t},
\]%
we note that in front of a test function%
\begin{eqnarray*}
\frac{\partial }{\partial t}\int \omega \varphi &=&\frac{\partial }{\partial t}%
\int \varphi (x_{0}(s,t))\mathbf{t=}\int \left( \left( v\cdot \nabla \varphi
\right) \mathbf{t}+\varphi \frac{d\mathbf{t}}{dt}\right) ds,\\
\int \left( v\cdot \nabla \omega \right) \varphi  &=&-\int \omega \left(
v\cdot \nabla \varphi \right) =-\int \left( v\cdot \nabla \varphi \right)
\mathbf{t}ds ,\\
\int \left( \omega \cdot \nabla v\right)  &=&\int \delta _{\Gamma }\frac{dv}{%
ds}\varphi =\int \frac{dv}{ds}\varphi (x_{0}(s,t))ds,
\end{eqnarray*}%
so that a singular filament is a formal solution if we choose the velocity
satisfying \eqref{c1}. Traditionally, the binormal law for the velocity is
obtained from Biot-Savart law under local induction approximation. We are
going to follow a different approach similar to the one in GP. Note first
that%
\[
v\cdot \nabla \omega -\omega \cdot \nabla v=\nabla \times (v\times \omega )
\]%
where, in each cross-section of the filament we can write in terms of the
stream function $\psi $
\begin{eqnarray*}
v &=&\nabla ^{\perp }\psi,  \\
\Delta \psi  &=&\left\vert \omega \right\vert.
\end{eqnarray*}%
Here $\nabla ^{\perp }\equiv(-\partial_y, \partial_x)$ denotes the skew gradient, so that%
\[
v\times \omega =\Delta \psi \nabla \psi
\]%
and we can follow the same steps as in \eqref{a2}--\eqref{biflow}.

\section{Vortex rings dynamics}
We are going examine the case of the dynamics of a vortex ring.  Early work of Levi-Civita \cite{LC} showed that small perturbations of a vortex ring under the binormal flow dynamics
undergo oscillatory motion. The formulation of previous section allow us to predict the behaviour of vortex rings evolution under the GP equation.

The analysis is easier when taking the formulation in terms of curvature and torsion. We will outline the derivation. By taking $dt' =- \left\vert \log \epsilon \right\vert  dt$ from \eqref{tgvel} and \eqref{biflow} we get
\begin{equation}
  \frac{d{\bf t}}{dt'} = \frac{d(\kappa {\bf b})}{ds}
\label{tt}
\end{equation}
where $s$ is the the arc-length parameter. Using the Frenet-Serret relations we can write \eqref{tt} as

\[
  {\bf t}_{t'}=-\kappa \tau{\bf n}+\kappa_s {\bf b}.
\]
Here $\tau$ denotes the torsion and the subindex the respective time and arc-length derivatives. We can use this expression to calculate ${\bf t}_{t's}$. On the other hand we have ${\bf t}_{s}=\kappa {\bf n}$, so taking the time derivative and equalling both expression we can solve for the time derivative of the normal vector to get
\[
\kappa\, {\bf n}_{t'}= \kappa^2 \tau {\bf t} - (\kappa_{t'} + 2\kappa_s \tau + \kappa\tau_s){\bf n}+(\kappa_{ss}-\kappa\tau^2){\bf b}.
\]
As the normal vector has a constant norm, we get our first equation for the evolution of the curvature

\[
\kappa_{t'} + 2\kappa_s \tau + \kappa\tau_s=0.
\]
To get the evolution of the torsion, we compute ${\bf n}_{t's}$ from the previous expression and the time derivative of ${\bf n}_{s}=-\kappa{\bf t} + \tau {\bf b}$, and equalling both expressions we can get ${\bf b}_{t'}$. Again, imposing that the ${\bf b}$ component is equal zero (and using $(|{\bf b}|^2)_t=0$) we get
\[
 -\tau _{t^{\prime }} +\left( \frac{\kappa _{ss}}{
\kappa }-\tau ^{2}\right) _{s}+\kappa \kappa _{s}=0.
\]
Rearranging the terms the equations for the  curvature and torsion reads,%
\begin{equation*}
\begin{split}
\kappa _{t^{\prime }} =& -\kappa \tau
_{s}-2\kappa _{s}\tau ,   \\
\tau _{t^{\prime }} =& \left( \frac{\kappa _{ss}}{%
\kappa }-\tau ^{2}\right) _{s}+\kappa \kappa _{s} ,  \label{eqt}
\end{split}
\end{equation*}

Around a vortex
ring of radius $R$ the equations can be linearised so that%
\begin{eqnarray*}
\kappa _{t^{\prime }} &\simeq &-\frac{\tau _{s}}{R}, \\
\tau _{t^{\prime }} &\simeq & R\kappa _{sss}+\frac{%
1}{R}\kappa _{s} ,
\end{eqnarray*}%
which can be combined to provide the evolution of the curvature,
\begin{equation*}
\kappa _{t^{\prime }t^{\prime }}=- \kappa _{ssss}-\frac{1}{R^{2}}\kappa _{ss} .
\label{eq:curvature}
\end{equation*}
The equation has a wave like structure, thus we can find solutions  of the form
\begin{equation}
\kappa (s,t^{\prime })=e^{i\frac{t^{\prime }}{R^2}\sqrt{n^{4}-n^{2}}}\cos \left( \frac{ns}{R}\right).
\label{eq:sol}
\end{equation}
The cases $n=0$ and $n=1$ are not very interesting as they correspond to a dilation and a rigid translation, respectively. However for $n\ge2$, the small perturbations will oscillate in time with a whole period of oscillation
\begin{equation*}
T=\frac{T'}{|\log \epsilon|}=\frac{2\pi R^2}{|\log\epsilon|\sqrt{n^4-n^2}}.
\end{equation*}

\section{Vortex rings oscillations}
In this section we are going to compare the time evolution of a vortex ring given by the GP equation with the one given by the binormal flow. We will solve GP numerically by considering the equation as a system for the real and imaginary parts of $u$, and implementing a finite elements method. This is in contrast with other numerical approaches such as in \cite{VKPS} based on tracking the vortex core. We first simulate the dynamics of a vortex ring of elliptical shape. We take as the initial condition for \eqref{eq:GP}, $u({\bf r}, 0) = u_0(r,\theta)\exp{i\theta}$, with

\begin{equation*}
  \label{eq:initvortex}
\begin{split}
  u_0(r,\theta)=&\left(1-e^{-\frac{(r-p)^2+(z-z_0)^2}{\epsilon^2}} \right)\left(1-e^{-\frac{(r+p)^2+(z-z_0)^2}{\epsilon^2}} \right),\\
  \theta=&\arctan\left( \frac{z-z_0}{r-p}\right)-\arctan\left( \frac{z-z_0}{r+p}\right),
\end{split}
\end{equation*}
being
\[
  r =\sqrt{ax^2+by^2}.
  \]
  \\
  \noindent
  The $z_0$ parameter controls the initial position along the vertical axis,
while $a$, $b$ and $p$ controls the ellipse parametrization. The eccentricity is given by $e = \sqrt{1 - b^2 /a^2}$.

\begin{figure}
  \centering
  \includegraphics[width=0.6\textwidth]{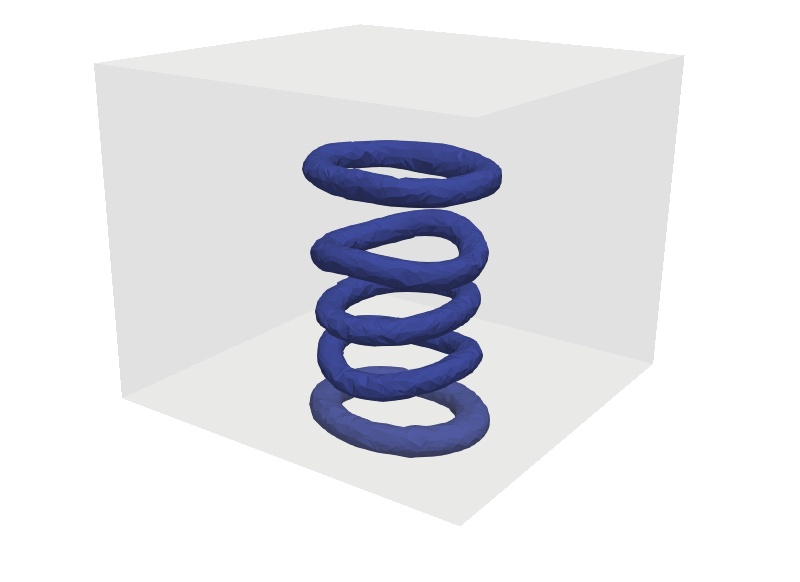}
  \caption{The evolution of a vortex of initial elliptical shape of eccentricity 0.6 using the GP equation. A periodic oscillation can be observed as the vortex moves in the vertical direction. We plot the level surfaces $|u|^2=0.3$.}
  \label{GPfig}
\end{figure}

According to the predictions of the previous section, for small eccentricities,  the elliptical case corresponds to  $n=2$ in \eqref{eq:sol}, so we expect the vortex ring will oscillate during its motion under the GP equation. In Fig.~\ref{GPfig} we simulate the case $a=1, p=0.5$, $\epsilon=0.1/\sqrt 2$ and $e=0.6$. We can see clearly that the vortex ring, approximately,  recovers the original shape after a time period.

\begin{figure}
  \centering
\includegraphics[width=0.9\textwidth]{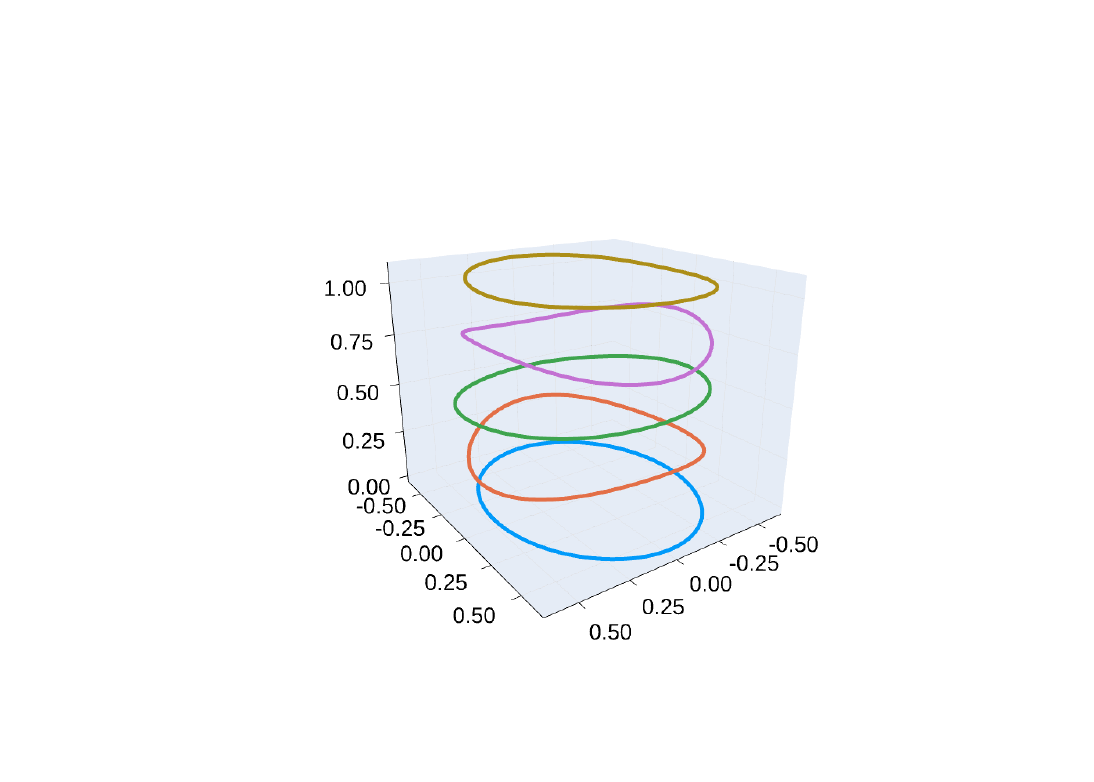}
  \caption{The evolution of a vortex of initial elliptical shape of eccentricity $0.6$ by the binormal flow. A periodic oscillation can be observed as the vortex moves in the vertical direction.}
  \label{binormalflow}
\end{figure}

\begin{figure}
  \centering
  \includegraphics[width=0.6\textwidth]{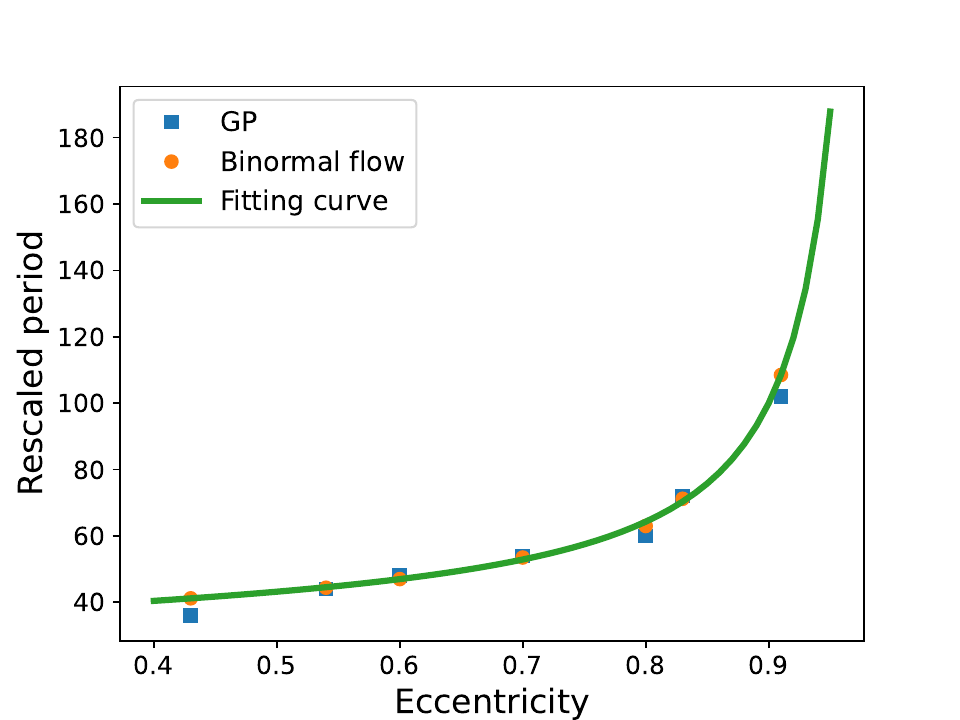}
  \caption{Dependence of the oscillatory recaled period $P=T/\epsilon^2|\log \epsilon|$ with the eccentricity of the vortex ring. Squares are computed using the GP equation, circles using the binormal flow. The fitting curve was calculated using the binormal points and a rational polynomial as explained in the main text.}
\label{periode}
\end{figure}

Now we can compare the GP dynamics with the binormal flow evolution characterised by \eqref{biflow}. Starting with an elliptical curve of the same eccentricity $e = 0.6$, we plot  its evolution under the binormal flow in Fig.~\ref{binormalflow}. Again the same oscillatory behaviour can be observed.

From these simulations we can study the dependence of the oscillatory period with the eccentricity. A larger eccentricity is expected to result in a longer period. We have done that using both GP and the binormal flow approximation, and the results show a satisfactory agreement as we can see depicted in Fig.~\ref{periode}. The solid line is the fitting of the observed rescaled period $P$ in the binormal flow using a rational polynomial. The rescaled period $P$ and the period $T$ are related by
\begin{equation}\label{TTT}
T(x)=\epsilon^2|\log \epsilon|P(x),
\end{equation}
where $x$ is the eccentricity of the initial data and
\begin{equation*}
  \label{eq:poly}
  P(x)=\frac{a_0+a_1x+a_2x^2}{1+b_1x}.
\end{equation*}
The optimal fitting coefficients for our data turn out to be
\begin{equation*}
  \label{eq:coef}
  a_0=33.45,\;a_1=-20.36, \; a_2=-7.71,\; b_1=-1.01.
\end{equation*}
If we compute the period for very small perturbations provided by formula \eqref{TTT} we have $P(0)=34.23$, which is very close to $a_0$.

\section{Vortex reconnections and self-similarity}
\begin{figure}
  \centering
  \includegraphics[width=0.4\textwidth]{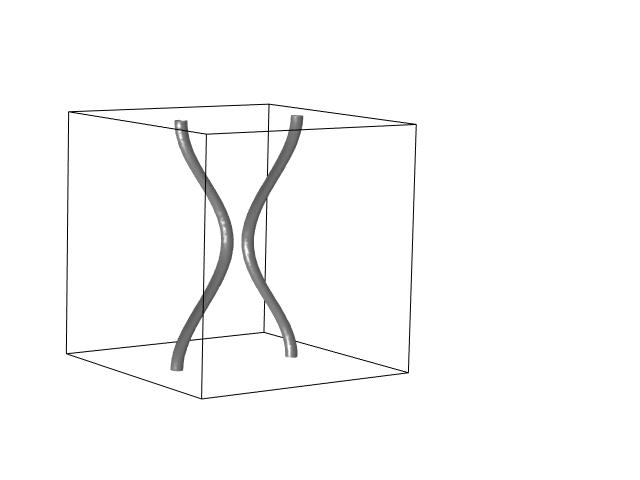}
  \includegraphics[width=0.4\textwidth]{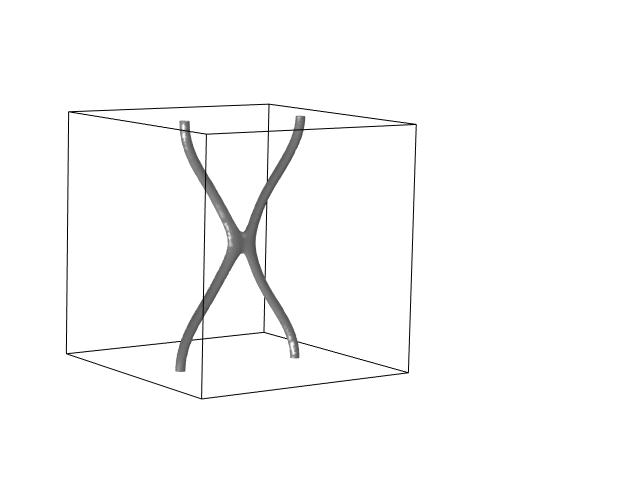}\\
  \includegraphics[width=0.4\textwidth]{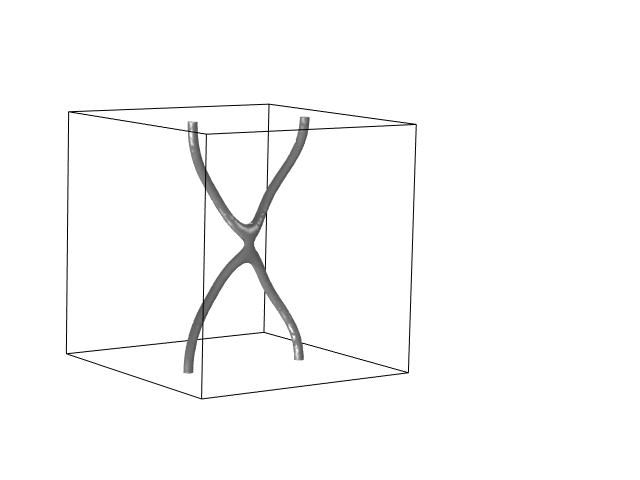}
  \includegraphics[width=0.4\textwidth]{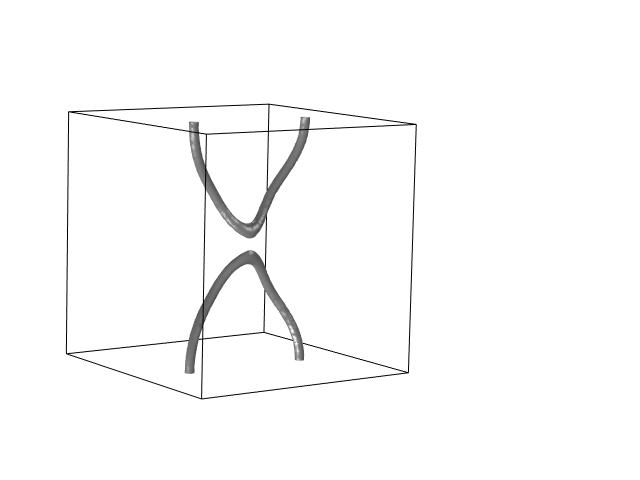}
  \caption{Snapshots of the vortex reconnection process in GP at different instants of time. From left to right and top to bottom, t = 0, 0.04, 0.1, 0.11.}
\label{vr}
\end{figure}

Now let us proceed to study the process of vortex reconnections. In Fig.~\ref{vr} we can see
the reconnection of two vortex filaments. They correspond to the simulations of the
GP equation using as initial conditions, $u(r, 0) = u_0 \exp {i\theta}$, with
\begin{equation*}
  \label{eq:inifil}
\begin{split}
   u_0=&\left(1-e^{-\frac{[x-x_0-p\cos(2\pi z/3)]^2+y^2}{\epsilon^2}} \right)\left(1-e^ {-\frac{[x+x_0+p\cos(2\pi z/3)]^2+y^2}{\epsilon^2}}\right),\\
  \theta=&\arctan\left( \frac{y}{x-x_0-p\cos(2\pi z/3}\right)-\arctan\left( \frac{y}{x+x_0+p\cos(2\pi z/3}\right),
\end{split}
\end{equation*}
and assuming periodic boundary conditions in the vertical direction. The values for
the parameters of the simulations shown in Fig.~\ref{vr} are $x_0 = 0.5$, $p = -0.35$ and $\epsilon = 0.1/\sqrt 2$.

\begin{figure}
  \centering
  \includegraphics[width=0.6\textwidth]{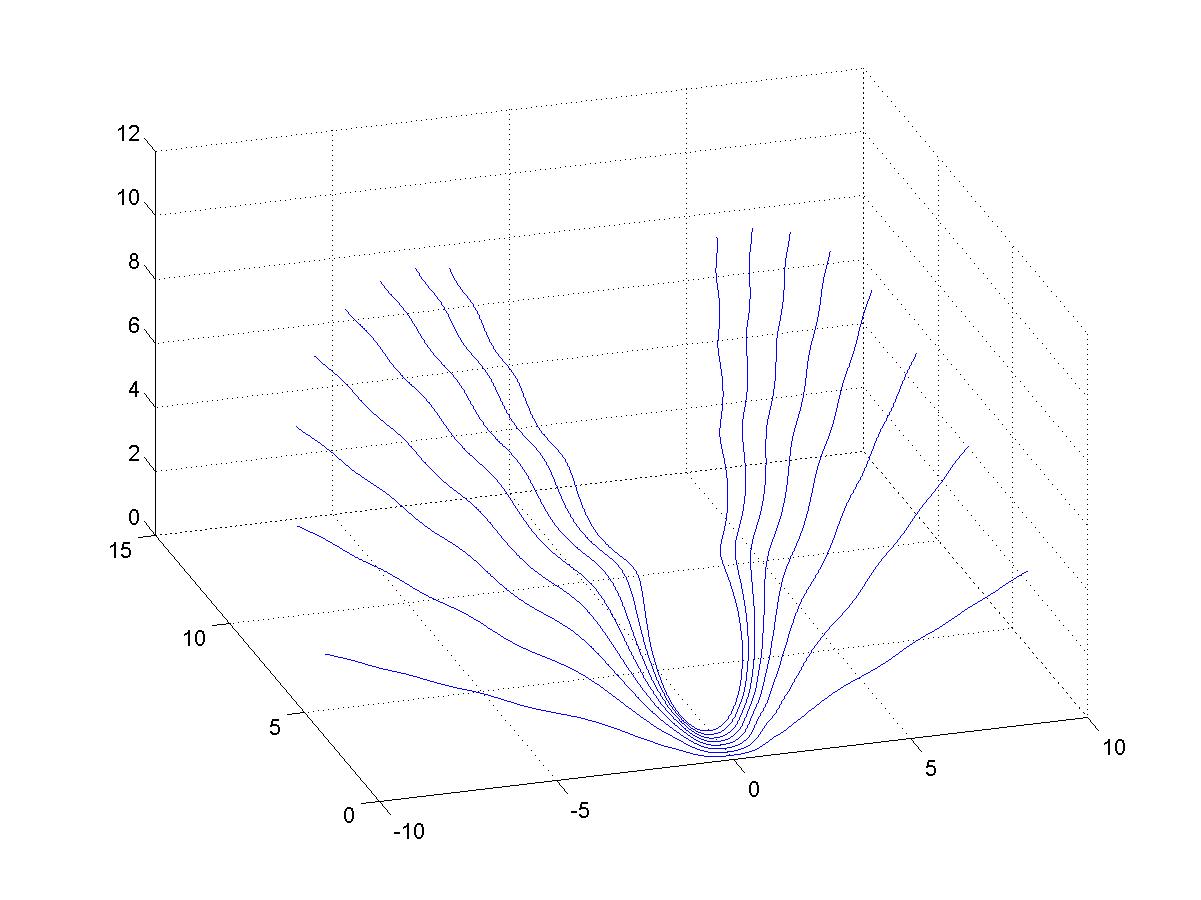}
  \caption{Self-similar solutions to the binormal flow equation forming a corner at $t\rightarrow 0$. Each plot corresponds to different initial opening angles.}
\label{auto}
\end{figure}

We observe that the geometry of the filaments is self-similar before and after the
reconnection. Let us find self-similar solutions to the binormal flow equation
\begin{equation*}
  \label{eq:bflow}
  {\bf x}_t=\kappa\, {\bf b}.
\end{equation*}
We consider solutions of the form
\begin{equation*}
  \label{eq:xself}
  {\bf x}(s,t)=t^{1/2}{\bf X}\left(\frac{s}{t^{1/2}}\right),
\end{equation*}
assume that reconnection happens at $t=0$ and look at filament
profiles after reconnection. By calling $\xi =s/t^{\frac{1}{2}}$ we find the
following equation for $\mathbf{X}$:%
\begin{equation}
\frac{1}{2}\mathbf{X-}\frac{1}{2}\xi \mathbf{X}_{\xi }=K\mathbf{b},
\label{aa1}
\end{equation}%
where $\kappa (s,t)=t^{-\frac{1}{2}}K(\xi )$. If we take the $\xi $
derivative of \eqref{aa1} we obtain%
\[
\mathbf{-}\frac{1}{2}\xi \mathbf{X}_{\xi \xi }=K_{\xi }\mathbf{b}+K\mathbf{b}%
_{\xi },
\]%
and using Frenet-Serret equations and the fact that $\mathbf{t}_{\xi }=%
\mathbf{X}_{\xi \xi }=K\mathbf{n}$ we deduce%
\[
\mathbf{-}\frac{1}{2}\xi K\mathbf{n}=K_{\xi }\mathbf{b}-K\tau \mathbf{n},
\]%
where $\tau $ is the torsion of the curve $\mathbf{X}(\xi )$. Hence,
projecting the last equation over the normal and binormal directions we get%
\begin{eqnarray*}
K_{\xi } &=&0, \\
\tau  &=&\frac{1}{2}\xi K,
\end{eqnarray*}%
i.e. a curve with constant curvature and linear torsion. We can numerically
compute a family of solutions parametrized by the curvature at $\xi =0$ and
whose tangent vector tends to be constant (say $\mathbf{T}_{\pm }$) as $\xi
\rightarrow \pm \infty $ (see Fig.~\ref{auto}), see also \cite{GRV} for the full analysis.

\begin{figure}
  \centering
  \includegraphics[width=0.45\textwidth]{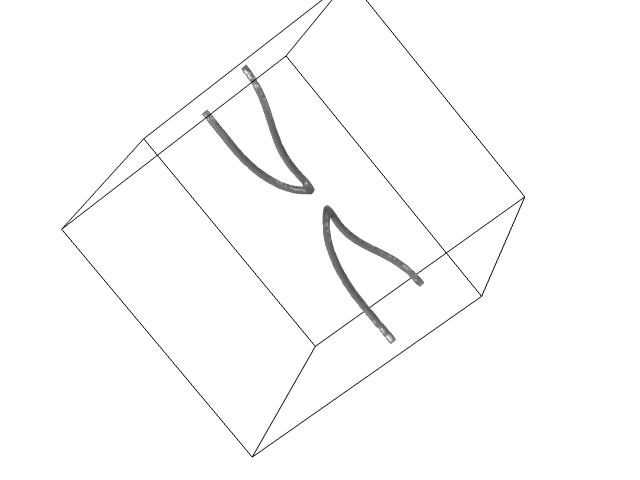}
  \includegraphics[width=0.45\textwidth]{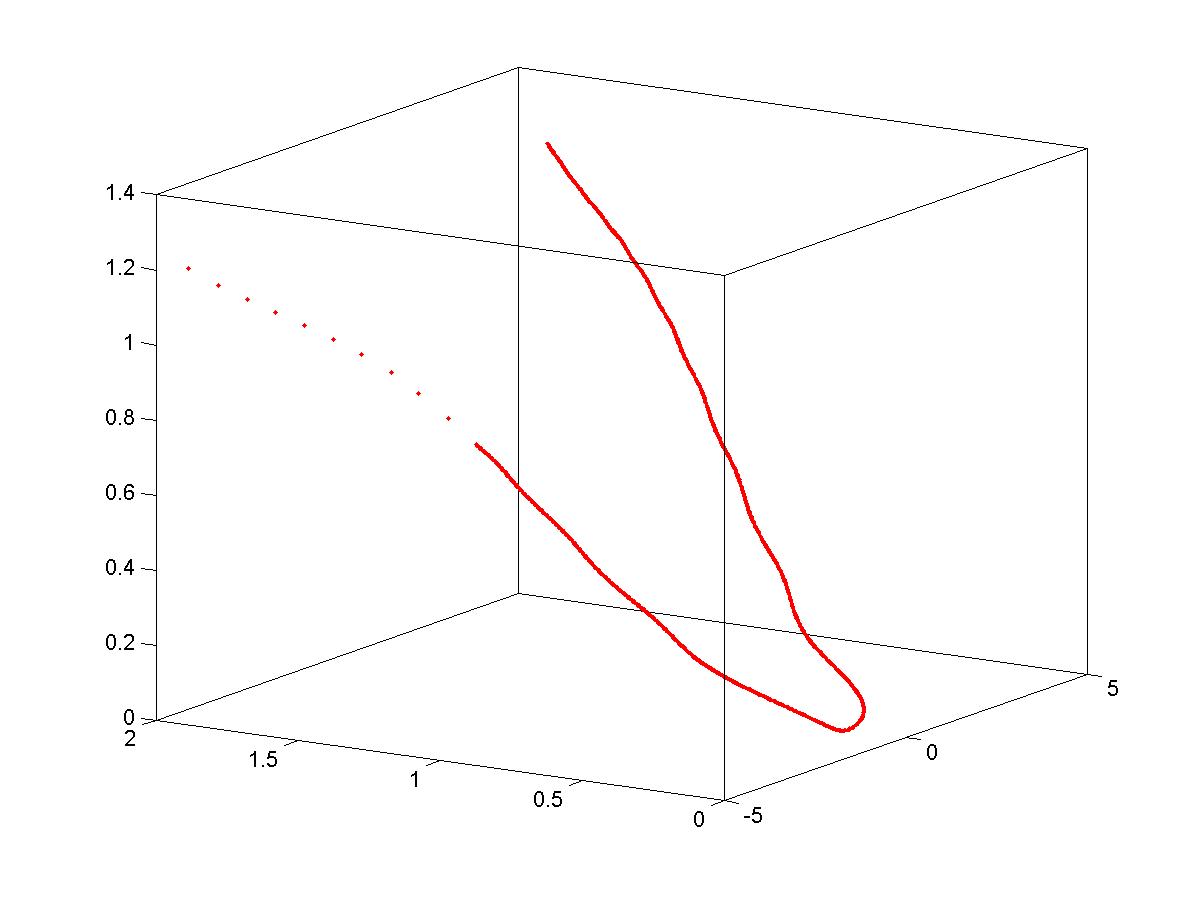}
  \caption{Left: close caption of the vortex filaments after reconnection by solving the GP. Right: the self-similar solution to binormal flow with the same opening angle and a similar orientation for comparison.}
\label{auto2}
\end{figure}

Note that, as $t\rightarrow 0$ the
curve $t^{\frac{1}{2}}\mathbf{X}(s/t^{\frac{1}{2}})$ tends to corner spanned
by those tangent vectors $\mathbf{T}_{\pm }$. Hence, after reconnection of
filaments in Gross-Pitaevski, and depending on the angle of the filaments at
reconnection, one could expect a self-similar evolution of the resulting
filaments in the form described above. Indeed, this is the case as shown in
Fig.~\ref{auto2}.

\section{Conclusions}

We have shown that, at leading order in the healing length $\epsilon$, Gross-Pitaevski equation has solutions in the form of vortex filaments that move following the binormal flow equation. In order to achieve this, we have reformulated the Gross-Pitaevski equation in a weak integral form that shares leading order terms (in $\epsilon$) to a weak formulation to the Euler system for inviscid fluids. This allows to deduce the precise form of the evolution equation for the vortex core.

    In addition, two numerical studies have been presented in order to demonstrate the approximantion by binormal flow. First, by taking as initial data for Gross-Pitaevski equation a vortex filament with the shape of a deformed vortex ring, we observe that the solutions computed numerically undergo oscillations with precise periods. We take the core of the initial vortex filament as initial data for the binormal flow and observe oscillations with almost the exact same periods ad Gross-Pitaevski solutions. Secondly, we observe the reconnection process by taking as initial data two almost parallel periodic filaments and observe that the geometry shorly before and after reconnection is self-similar with similarity laws and profiles almost identical to the theoretical ones for binormal flow equation.

\section*{Acknowledgments}
This work was supported by Grants PID2022-139524NB-I00 and PID2020-113596GB-I00, funded by MCIN/AEI/10.13039/501100011033.


\begin{thebibliography}{123456}% common bib file

\bibitem{Vinen02} W.~F.~Vinen, J.~J.~Niemela, Quantum turbulence,  J. Low Temp. Phys.
  {\bf 128}, 167–231 (2002).

\bibitem{Batchelor} G.~K.~Batchelor,  An Introduction to Fluid Dynamics, Cambridge Univer-
sity Press, 1967.

\bibitem{Schwarz85} K.~W.~Schwarz,  Three-dimensional vortex dynamics in superfluid $^4$He: Line-line and line-bounday interactions, Phys. Rev. B {\bf  31} 5782 (1985).

\bibitem{Schwarz88} K.~W.~Schwarz,  Three-dimensional vortex dynamics in superfluid $^4$He: Homogeneous superfluid turbulence, Phys. Rev. B {\bf  38} 2398 (1988).

\bibitem{Adachi10} H. Adachi, S. Fujiyama, M. Tsubota, Steady-state counterflow quantum turbulence: Simulation of vortex filaments using the full Biot-Savart law.  Phys. Rev. B {\bf  81} 104511 (2010).

\bibitem{Gross61} E.~P.~Gross,  Structure of a quantized vortex in boson systems. Il Nuovo Cimento. {\bf 20 (3)} 454--457, (1961).

\bibitem{Pitaevskii61} P.~Pitaevskii, Vortex lines in an imperfect Bose gas. Sov. Phys. JETP. {\bf 13} (2): 451--454 (1961).

\bibitem{Syvokon} V.~E.~Syvokon, Influence of the superfluid transition on the adsorption of thin helium films. Low Temp. Phys. {\bf 32} 48 (2006).

\bibitem{LC} T.~Levi-Civita, Teoremi di unicità e di esistenza per le piccole oscillazioni
di un ﬁletto vorticoso, prossimo alla forma circolare. R. C. Accad. Lincei (6) {\bf 15}, 409 (1932).

\bibitem{ATN} T. Araki, M. Tsubota, and S. K. Nemirovskii, Energy Spectrum of Superfluid Turbulence with No Normal-Fluid Component.
Phys. Rev. Lett. \textbf{89} (2002), 145301.

\bibitem{VKPS}
A. Villois, G. Krstulovic, D. Proment and H. Salman, A vortex filament tracking method for the Gross–Pitaevskii model of a superfluid. J. Phys. A: Math. Theor. \textbf{49} (2016), 415502.

\bibitem{GRV} S. Guti\'errez, J. Rivas and L. Vega, Formation of singularities and self-similar vortex motion under the localized induction approximation, Commun. PDE \textbf{28} (2003) 927--968.

\bibitem{Jerard} R. Jerrard. Vortex filament dynamics for Gross-Pitaevsky type equations. Annali della Scuola Normale Superiore di Pisa - Classe di Scienze, Serie 5, Volume \textbf{1} (2002) no. 4, pp. 733--768.


\end{thebibliography}
\end{document}